\documentclass[aps,prl,twocolumn,showpacs,superscriptaddress,groupedaddress]{revtex4}
\usepackage{graphicx}  
\usepackage{dcolumn}   
\usepackage{bm}        
\usepackage{amssymb}   
\usepackage{amsmath}

\begin{document}



\title{Coulomb driven energy boost of heavy ions for laser plasma acceleration}
\author{J. Braenzel}
\affiliation{Max Born Institute, Max Born Str. 2A, 12489 Berlin,
Germany} \affiliation{Technical University Berlin, Strasse des 17.
Juni 135, 10623 Berlin, Germany}

\author{A. A. Andreev}
\affiliation{Max Born Institute, Max Born Str. 2A, 12489 Berlin,
Germany} \affiliation{Vavilov State Optical Institute, Birzhevaya
line 12, 199064 St. Petersburg, Russia} \affiliation{St. Petersburg
University, University emb.6, St. Petersburg 199064, Russia}

\author{K. Platonov}
\affiliation{Vavilov State Optical Institute, Birzhevaya line 12,
199064 St. Petersburg, Russia}

\author{M. Klingsporn}
\affiliation{IHP, Im Technologiepark 25, 15236 Frankfurt}

\author{L. Ehrentraut}
\affiliation{Max Born Institute, Max Born Str. 2A, 12489 Berlin,
Germany}

\author{W. Sandner}
\affiliation{Max Born Institute, Max Born Str. 2A, 12489 Berlin,
Germany} \affiliation{Technical University Berlin, Strasse des 17.
Juni 135, 10623 Berlin, Germany} \affiliation{ELI-DC International
Association AISBL}

\author{M. Schn\"{u}rer}
\affiliation{Max Born Institute, Max Born Str. 2A, 12489 Berlin,
Germany}

\date{\today}

\begin{abstract}
An unprecedented increase of kinetic energy of laser accelerated
heavy ions is demonstrated. Ultra thin gold foils have been
irradiated by an ultra short laser pulse at an intensity of $6\times
10^{19}$ W/cm$^{2}$. Highly charged gold ions with kinetic energies
up to $> 200$ MeV and a bandwidth limited energy distribution have
been reached by using $1.3$ Joule laser energy on target. $1$D and
$2$D Particle in Cell simulations show how a spatial dependence on
the ions ionization leads to an enhancement of the accelerating
electrical field. Our theoretical model considers a varying charge
density along the target normal and is capable of explaining the
energy boost of highly charged ions, leading to a higher efficiency
in laser acceleration of heavy ions.

\end{abstract}

\maketitle


Laser driven ion acceleration has gained a wide scientific interest,
as it is a promising ion source for investigation in basic plasma
physics and for application in accelerator technology
\cite{Cowan2004.01,Scuderi2014.01} related to bio-medical
\cite{Schardt2010.01,Tajima2009.01} and hadron research \cite{CERN}.
While the acceleration of protons and light ions are intensively
investigated during the last decade, little is reported on
acceleration of heavier ions \cite{Hegelich2002.01}. Such knowledge
is mandatory to achieve the objectives of upcoming new laser
facilities \cite{ELI,daido}, e.g. the exploration of nuclear,
astrophysical questions as well as the potential use as beam lines
for heavy ion radio therapy \cite{Kraft}.

Energies of heavy ions exceeding the mass number $A\gg12$ with
$E_{kin}/{u}\sim 1-2$ MeV/u (energy per nucleon) have been reported
so far \cite{Hegelich2002.01,Hegelich2006.01}, by using short pulse
laser systems with laser pulse energies well above $20$ J
\cite{Hegelich2005.01}.

In the following we report and discuss a considerable energy boost
for acceleration of the highly charged heavy ions with only using
$1.3$ J on an ultra thin heavy material target. We accelerated ions
up to $E_{Max}/{u}>1$ MeV/u, with a bandwidth limited energy
distribution. We found a remarkable deviation in the maximum energy
to charge $Z$ scaling in comparison to established models of Mora
 \cite{Mora2005.01} and Schreiber \cite{Schreiber2006.01,Schreiber}.


Presently used laser ion acceleration schemes like Target Normal
Sheath Acceleration (TNSA) \cite{wilks2001.1}, or leaky light sail /
Radiation Pressure Acceleration (RPA) \cite{Esirkepov2006.01,
Qiao2010.01,Henig2009.02}, Coherent Acceleration of Ions by Laser
(CAIL)  \cite{Yan2008.01,Tajima2009.01}, Break Out Afterburner (BOA)
\cite{Jung2013.01} make use of an energy transfer from laser to
electrons and in a following step electrons accelerate the ions. In
the typical physical picture, an ultra intense laser is focused on a
thin target, ionizes it and displaces the electrons from the ion
background by the laser field. This creates a high electrical field
at the rear and front side of the target. The Coulomb attraction
field of the ions circumvents the electrons escape and enables the
acceleration of the ions. For ultra thin targets and relativistic
laser intensities, the acceleration is enhanced by the transparency
of the target and the relativistic kinematics of the electrons
\cite{Yan2010.01,Henig2009.02,Steinke2010.01,Steinke2013.01}.
Further optimization for the energies of light ions is proposed by a
Coulomb exploding background of heavy ion constituents in a ultra
thin foil target \cite{Esirkepov2002.01, Qiao2012.01,
Bulanov2008.01}. A remarkable contribution by the Coulomb explosion
to the energy of very heavy ions energy is predicted but still under
theoretical discussion \cite{Wang2014.01, Korzhimanov2012.01}.

Most acceleration models assume an averaged degree of ionization
leading to a fixed electron density - which creates the moving
accelerating electrical field for the ions. During the laser plasma
interaction ions of different charge to mass ratio $Z/A$ separate in
the velocity picture, leading to higher MeV/u for the lighter
material. The energy per nucleon decreases significantly with the
decreasing charge to mass ratio, as the accelerating field is
screened by the light ions. Laser plasma experiments using thin
foils showed, that in the presence of hydrogen and carbon, ions with
a smaller $Z/A$ ratio are not accelerated at all or stay with much
lower velocity \cite{Hegelich2006.01}. Only specially prepared,
heated targets without contamination by light ions, enabled an
acceleration of the heavy ions up to the MeV/u range. To our
knowledge we obtained for the first time heavy ions with $>1$ MeV/u
in presence of the contamination layer. While the maximum kinetic
energy $E_{kin}^{Max}$ for hydrogen reach $12$ MeV/u and $4.2$ MeV/u
for $C^{6+}/O^{8+}$, the highest charged gold ion $\gtrsim Au^{50+}$
follows with $\gtrsim 1$ MeV/u.

\begin{figure}
\includegraphics[scale=0.5]{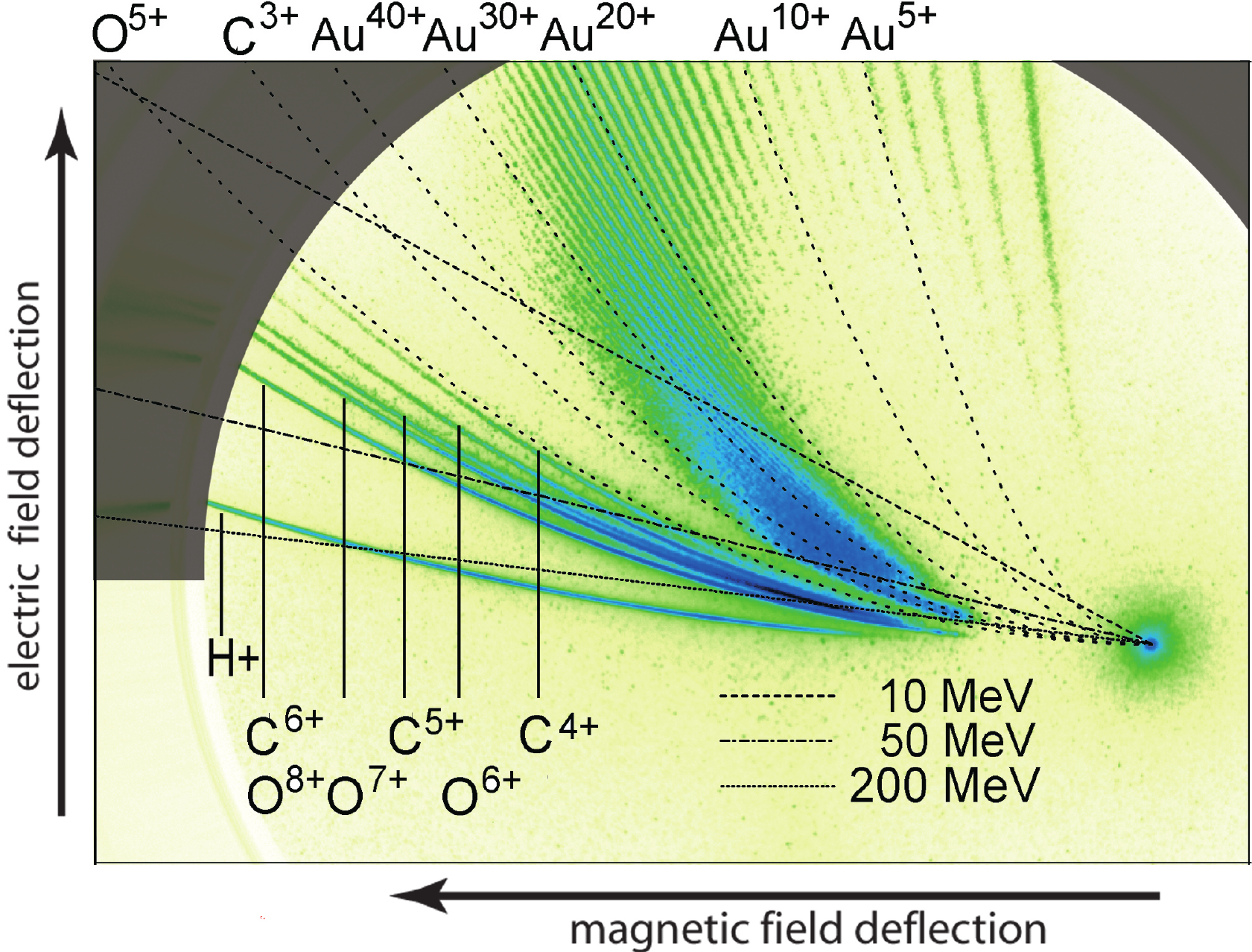}
\caption{\label{Fig1} Raw spectra from Thomson spectrometer (single
shot measurement), particle density in false color coding. Each
trace represents a different charge to mass ratio $Z/m$. Gray shade
indicates end of detector. Light ion traces ($H^{+}$,
$C^{6+}-C^{3+}$,$O^{8+}-O^{5+}$) are identified. Overlay shows
theoretical parabolas at different charge states of gold ions (black
dots). Straight lines mark theoretical constant energy at each
degree of ionization for gold, $m=197$ u.}
\end{figure}

Experiments have been performed at the Max Born Institute High Field
Ti:Sapph. laser. It delivers $1.3$ J at $(30-35)$ fs on the target
after contrast enhancement by a XPW \cite{Klashnikov2011.01}
frontend and a Double Plasma Mirror (DPM) \cite{Levy2007.01},
leading to a pre pulse free peak to ASE contrast of $\leq10^{-14}$
in the minor ps range. The laser is focused by a $f/2.5$ off axis
parabola to a focal FHWM size of $\sim 4$ $\mu$m, giving an averaged
intensity of $6 \times10^{19}$ W/cm$^{2}$ in the focal area. The
normalized laser field is $a_{0}=qE_{L}/m_{e}c\omega =5$ for linear
polarization, with the electron mass $m_{e}$ and charge $q$, laser
frequency $\omega$ and speed of light $c$, respectively. We focused
the laser at free standing $(14\pm2)$ nm gold foil
\cite{Braenzel2013.01}, which we produced by thermal evaporation at
$10^{-6}$ mbar (deposition rate: $0.2$ nm/s), followed by a floating
process. HRTEM (High Resolution Transmission Electron Microscopy)
reveals a polycrystaline structure of the gold formed by an island
growth mode on a carbon based supportive film, which we identify as
the rest of the parting agent. The average grain size is of the
order of $10$ nm. Determination of the composition has been carried
out by EDX (Energy Dispersive X-Ray Spectroscopy) with a state of
the art FEI ChemiSTEM\texttrademark system and was quantified
standardless with a Cliff-Lorimer calculation. The foil consists of
gold $~96\%$, carbon $2-3\%$ and oxygen $~2\%$, hydrogen is not
determinded. STEM (Scanning Transmission Electron Microscopy)
measurements reveals a sub crack like structure in ($10-20$) nm
distance (see Fig.\ref{Fig3}b). Structured surfaces can increase the
absorption of the laser light, leading to a higher efficiency of the
acceleration mechanism. This is discussed widely at the moment, but
yet has not been considered for the thinnest targets
\cite{Andreev2013.02,Ceccotti2013.01}. Accelerated particles were
detected in single shot measurement by a Thomson spectrometer at
$0^{\circ}$ in laser propagation direction. The setup consists of an
entering pinhole with a diameter of $110$ $\mu$m, a permanent
magnet, electrical field plates and a $100$ mm Multi Channel Plate
(MCP Hamamatsu) covering a detection angle of $1\times 10^{-7}$ sr
from the target  \cite{sokollik}. Measurements at a lower laser
contrast (without DPM) $\leqq10^{-11}$, showed much lower
$E_{kin}^{Max}$ and particle numbers for hydrogen, carbon, oxygen
ions and no gold ion spectra in the measured energy range.

\begin{figure}
\includegraphics[scale=0.3]{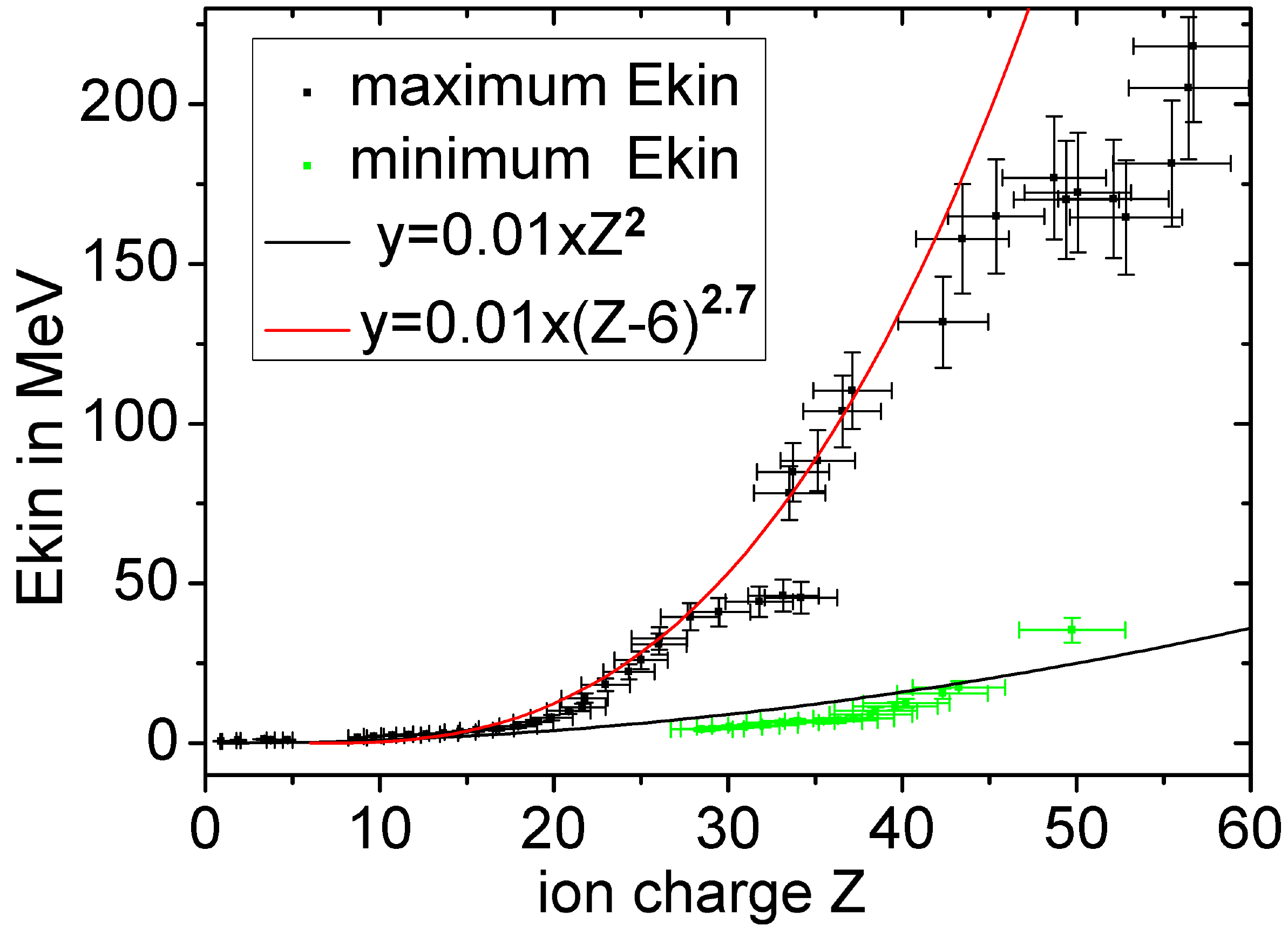}
\caption{\label{Fig2} Maximum (black) and minimum (green) kinetic
energy of gold ions in dependence of their charge state Z. For
$Z<25$ the detectors range is cutting the low energetic part of the
spectra. Red line shows a $(Z-6)^{2.7}$ and black line a $Z^{2}$ to
$E_{kin}^{Max}$, both fit functions with the same scaling
coefficient.}
\end{figure}

\begin{figure}
\includegraphics[scale=0.7]{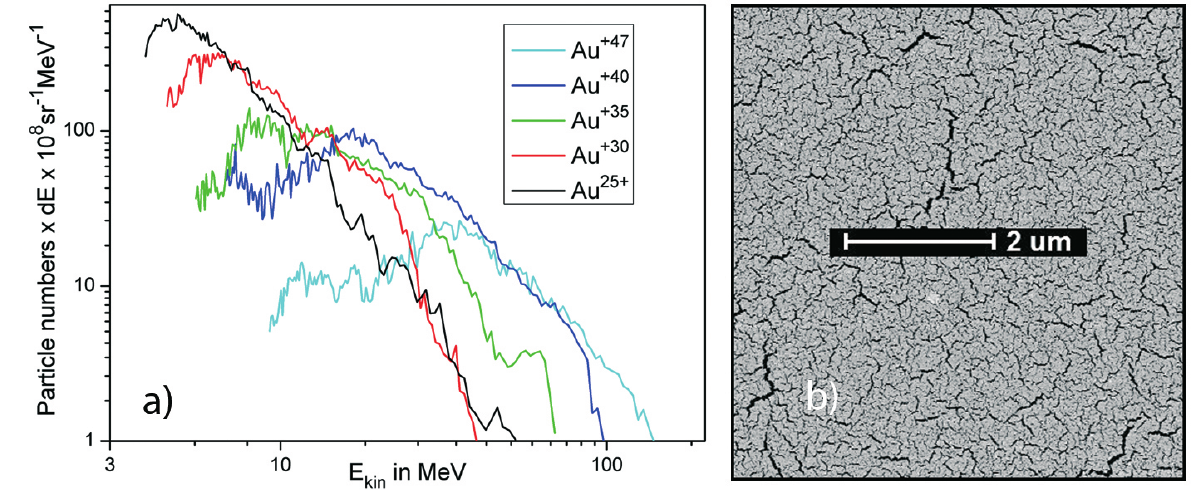}
\caption{\label{Fig3} \textbf{a:} The associated evaluated energy
distribution for selected, single traces of gold ions of
Fig.\ref{Fig1} and Fig.\ref{Fig2} is shown, exhibiting a pronounced
maximum. $dE$ is given by the binning of the spectrometers
resolution. \textbf{b:} STEM measurement of freestanding target foil
reveals a crack like structure. Dark cracks mark here the carbon
substrate layer.}
\end{figure}

Fig.\ref{Fig1} shows a captured picture of the detector. We identify
traces of accelerated gold particles for ionization degrees reaching
from $Au^{1+}$ to $>Au^{50+}$, well beyond the $C^{3+}$ trace. With
increasing charge to mass ratio we observe light ions traces of
oxygen, carbon and hydrogen. For a quick interpretation of the
measured data, the overlay in Fig.\ref{Fig1} shows lines of constant
energy for $m=197$ u to mark constant energy positions for different
charge states on the detector. We observed a strong signal for gold
ions between $Au^{20+}$ and the highest degree of ionization
$>Au^{50+}$ with kinetic energies from $10$ MeV to $200$ MeV. The
traces exhibit a distinct maximum in particle numbers and a
bandwidth limited energy distribution for charge states $Z>25$. The
low energetic cutoff for ions charged $Z<25$ probably lies beyond
the detection range. The symmetry of the gold ions cutoff on the
detector seems to follow a lemniscate like function (half figure
eight): $r(\phi(Z))\sim a^{2}\times 2\sin(2\phi(Z))$, with $a$ as a
constant of the radius and $\phi(Z)$ a nonlinear, charge depending
function. We evaluated the highest energy cutoff and lowest energy
cutoff for the different charge states of gold ions in
Fig.\ref{Fig2}.
Compared to a expected $E_{kin}^{Max}\propto Z^{2}$ scaling by the
model of \cite{Schreiber2006.01}, our data shows a boosted scaling
of $E_{kin}^{Max}\propto (Z-6)^{2.7}$. For a better comparison
Fig.\ref{Fig2} uses the same scaling coefficient for both fit
functions. Experiments with gold coated plastic foils (Formvar
($10-40$)nm $+$ ($2-6$)nm gold coating on target rear side) showed
similar results concerning the multiple degrees of ionization, the
$Z$ to $E_{kin}^{Max}$ scaling, reaching close to the MeV/u range
and with a limited bandwidth in the energy spectrum (see
supplement). It reveals a general mechanism for the acceleration of
heavy ions if ultra thin foil with heavy material is used. The
energy distribution related to Fig.\ref{Fig1} of selected gold ions
is shown in Fig.\ref{Fig3}. The particle numbers are given relative
to a detector calibration with hydrogen and carbon, assuming a
similar response for heavy ions \cite{Prasad2010}. We approximated
the energy content of all accelerated gold ions with
\cite{Nurnberg,Bambrink} (for methods see supplement) to $5\%$ of
the laser energy, while the $H^{+}$ reaches $<2\%$.

In order to account for the theoretical ionization $Z$ in dependence
on the electrical field strength $a_{0}$ we used the ADK model
\cite{Ammosov1986.01}. The calculation for gold is shown in
Fig.\ref{Fig3}a) and we find an ionization dependence
$Z(a_{0})=23\times a_{0}^{0.4}$. The field strength for our
parameters considers an intensity of $a_{0}=5$, which leads to a
maximum ionization of $Z(a_{0})=42$. Higher ionization as observed
in our experiment can be attributed to field enhancement in case of
partly transparent target plasma, to contributions from the surface
structure and to self focusing.

\begin{figure}
\includegraphics[scale=0.35]{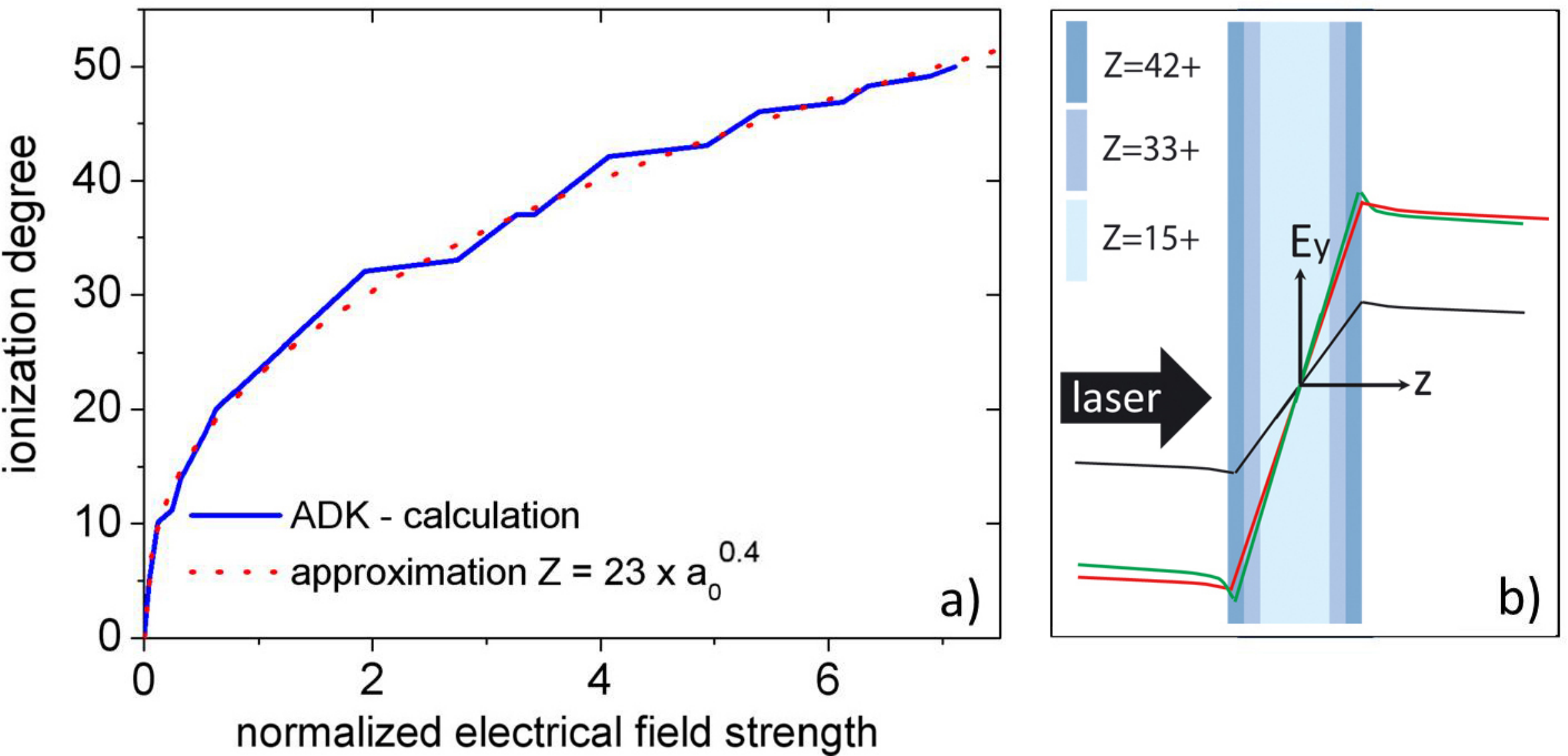}
\caption{\label{Fig4} \textbf{a:} The dependence of gold ionization
on the electric field $E_{L}$ in units of $a_{0}$ calculated with
the ADK-Model. The dashed line (red) fits $Z(a_{0})=23\times
a_{0}^{0.4}$. \textbf{b:} $E_{L}$ calculated from analytical model
(red) and PIC simulation (green) considering ion layers of the
following degrees of ionization: The distribution of ion charge is:
$0-1$nm $Z=42$, $1-2$nm $Z=33$, $2-18$nm $Z=15$, $18-19$nm $Z=33$,
$19-20$nm $Z=42$. Black line - $E_{L}$ calculated with an averaged
ionization degree of $Z=15$.}
\end{figure}

Our $1$D PIC simulation evaluated at high accuracy (mesh size:
$0.16$ nm, $200$ particles per cell, error $<1\%$) has been
performed using the laser parameters of the experiment and a target
thickness of $20$ nm. For simplification we freezed the ionization
in time at the end of the laser pulse. The $1$D PIC simulation shows
in longitudinal direction a symmetrical, varying ionization degree
$Z(z)$ (see Fig.\ref{Fig4}b) \cite{Zhidkov}. Compared to an averaged
degree of ionization, it leads to an enhancement of the electrical
field at the front and rear side of the target by contributions of
the repelling Coulomb force. The field enhancement becomes strong
for highly charged ions.

For the $2$D PIC simulation we used $5\times 10^{19}$ W/cm$^{2}$,
$35$ fs, $4$ $\mu$m focus diameter, Gaussian laser profile. The
pulse interacts with a pure $20$ nm thick gold target. The step size
of the calculation was $0.5$ nm with $30$ particles per cell. In
Fig.\ref{Fig5} we compare the calculated energies with our
experimental results and the model of \cite{Schreiber2006.01}. The
$E_{kin}^{Max}$ to $Z$ dependence has to be separated into three
parts: while for $Z<15$ the Au ion energies fit to a
$E_{kin}^{Max}\sim Z^{2}$, ions with $Z>15$ are with an exponent
$>2$, followed by a smaller linear dependence for $Z>42$.

Our analytical model focuses on the Poisson equation, as the
electrical field of the laser does not penetrate deep inside even in
our thin foil. We take a spatially varying ionization of heavy
target material into account:

\begin{equation}
2(\frac{\partial^{2}\eta_{e}}{\partial \xi^{2}}+
\frac{\partial^{2}\eta_{e}}{\partial \varsigma^{2}})=
\eta_{e}-Z(E)n_{i0}\Theta(\frac{l_{f}}{2}-|\xi|)
\Theta(\frac{l_{e}}{2}-|\varsigma|) \label{eq:one}
\end{equation}

Here we use a $2$D geometry with the coordinates
$(z,y)=(\xi,\varsigma)r_{D}$, where the Debye radius is
$r_{D}^{2}=T_{H}/4\pi e^{2}n_{eH}$ and assuming the process to be
adiabatic. The normalized electron density is
$\eta_{e}=n_{e}/n_{eH}=1+\phi /2$ and the normalized electric field
is $E=\frac{2c}{\omega r_{D}}\frac{\partial \eta_{e}}{\partial
\xi}$. The ion density
$n_{i}(z,y)=\eta_{i0}\Theta(z)\Theta(y)n_{eH}$ has a rectangular
shape in both directions, where $L_{f,e}(t)=l_{f,e}(t)r_{D}$ are
dynamic foil thickness and electron spot size, respectively. The hot
electron density is determined from quasi-neutrality and the
ionization degree is $Z(E)=23E^{0.4}$. We introduce a spatial
dependence of the hot electron density: $n_{eH}\approx \frac{\pi
e^{2} n_{i}^{2}}{T_{H}}(\int_{0}^{l_{f0}}Z(\xi))^{2}$. The spatial
depending degree of ionization is given as:

\begin{equation}
Z(\xi)=23\times(\frac{2c}{\omega r_{D}}\frac{\partial
\eta_{e}}{\partial\xi})^{0.4} \label{eq:two}
\end{equation}

The electron temperature $T_{H}$ depends on the pulse duration
$\tau_{L}$ and on a laser absorption coefficient $\kappa$ (here and
in the following see \cite{Andreev2010.01}):  $T_{H}(l_{f})\approx
\frac{\kappa(l_{f0})I_{L}\tau_{L}}{n_{eh}l_{f0}}$. For
simplification, we assume a rectangular transversal $(y)$ and
longitudinal $(z)$ electron density profile, which width changes in
time with $l_{f}(t)$. For the ultrathin foil follows:
$\Theta(l_{f}(t)/2-|\xi|)\rightarrow l_{f}(t)\delta(\xi)$, we take
the expansion of the recirculating hot electrons as a time dependent
parameter $l_{e}(t)$. At this point we freeze the degree of
ionization in time. The time dependent solution of Eq.\ref{eq:one}
at $|z|\geq l_{f}(t)/2$  looks similar to \cite{Andreev2010.01}:

\begin{figure}
\includegraphics[scale=0.25]{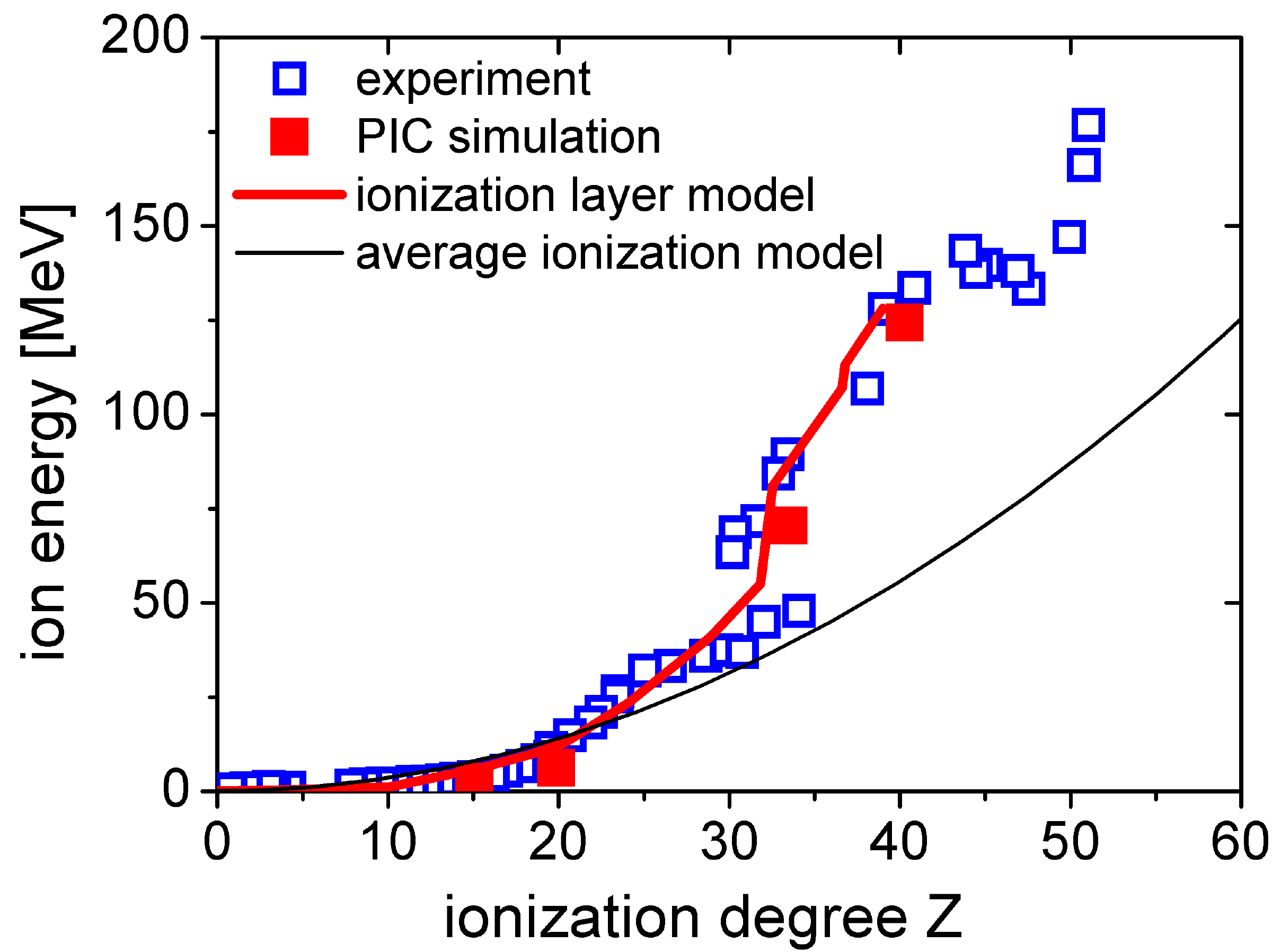}
\caption{\label{Fig5} The dependence of maximal ion energy on its
ionization degree: the experimental data of Fig.\ref{Fig2} - deep
blue squares, $2$D PIC - simulation data - red squares, Schreiber
model - black line and our model - red line. The distribution of ion
ionization is according to the $1$D PIC simulation in
Fig.\ref{Fig4}.}
\end{figure}

\begin{equation}
\begin{split}
E(z,y,t) & = 4\pi e
n_{i0}\frac{sign(z)\Theta(l_{e}(t)-|\varsigma|)}{1+\sigma_{c}tr_{D}l_{f}(t)/D_{0}^{2}} \\
& \times\int_{0}^{l_{f0}}Z(\xi)d\xi\exp(-|\xi|+\frac{l_{f0}}{2})
\label{eq:three}
\end{split}
\end{equation}

$D_{0}$ denotes the initial electron spot size and $\sigma_{c}$ is
the plasma conductivity. The equation contains a spatial dependence
of the charge distribution in the target instead of an averaged,
constant one. The dependence of the analytical field
(\ref{eq:three}) on coordinate $z$ is similar to the PIC simulated
one (Fig.\ref{Fig4}b). The charged ion front $l_{f}(t)$ in the
target can be calculated by the equation of motion after inserting
(\ref{eq:three}) and with $C=16\pi {e}^{2}l_{f0}n_{i0}/m_{i}$:

\begin{equation}
\left. l_{f}(t)= l_{f0}+t\sqrt{C\times
Z(l_{f0})[\int_{0}^{l_{f0}}Z(\xi)d\xi] \ln (\frac
{l_{f}(t)}{l_{f0}})}\right. \label{eq:four}
\end{equation}

Expression (\ref{eq:four}) defines the energy of an ion with maximum
degree of ionization, which is at the front of acceleration
$\varepsilon_{Z(l_{f0})}=m_{i}\dot{l}_{f}^{2}(t)/8$. Electron
density in each instant is defined by (\ref{eq:one}). From the
equation of continuity follows
$n_{i}(z,t)=n_{i}l_{f0}\Theta(l_{f}/2-|z|)/l_{f}(t)$ and the ion
velocity  with the coordinate of $z$ reads:
$v_{i}(z,t)=z\dot{l}_{f}(t)/l_{f}(t)$, $|z|<l_{f}(t)/2$. The energy
for a particle placed initially at $\xi_{0}$ with an charge of
$Z(\xi_{0})$ has to be evaluated parametrically with (\ref{eq:two})
and (\ref{eq:four}). For ions inside the target
$\xi_{0}\epsilon[0,l_{f}/2]$ results:

\begin{equation}
\left.{
\varepsilon_{z}(\xi,t^{\ast})=\frac{m_{i}}{2}(\xi_{0}/l_{f0})^{2}\dot{L}_{f}^{2}(t^{\ast})}
\right.{\label{eq:five}}
\end{equation}

With $t^{\ast}\approx D_{0}^{2}/\sigma_{c}r_{D}l_{f}$ for ions of
very high energy $t^{\ast}\sim2\tau_{L}$ \cite{Mora2005.01}. This
leads to $\sim Z^{3}$ ion energy to charge scaling, which is in good
agreement with our PIC simulated and experimental results (see
Fig.\ref{Fig5}).

Ions with a very high degree of ionization $Z>Z(l_{f0})$, are formed
in a field maximum at the target rear side. These ions have the
initial coordinate $\xi_{0}=l_{f0}$. According to (\ref{eq:four})
$\dot{l}\approx\sqrt{Z(l_{f0})})$ for ions with a high charge $Z$,
the formula (\ref{eq:five}) gives for all $Z>42$ the linear relation
$\varepsilon_{Z}\sim Z$. The smaller energy to $Z$ scaling is
explained by the decreasing charged background compared to ions
placed inside the target.\\

 In conclusion, we demonstrated efficient acceleration of heavy
ions by an ultra short laser pulse system. So far laser systems that
compensate lower laser energy with a shorter pulse duration to reach
the same intensity, had not been able to accelerate heavy ions with
$A>12$ into the MeV/u region. By using an ultra thin foil of heavy
material we achieved highly charged heavy ions with a limited
bandwidth in the energy spectrum, reaching up to $1$ MeV/u.
Furthermore we simplified a complex target preparation, which
achieves a prerequisite for future applications. We demonstrated
experimentally and theoretically how a spatial distribution of the
ionization inside the target leads to a field enhancement for the
heavy ions by Coulomb explosion. This has the potential to greatly
improve the efficiency of heavy ion acceleration by stronger kinetic
energy with charge scaling. Our results indicate that e.g. energies
with $7$ MeV/u can be achieved with $\sim50$ times higher laser
energy than in our experiment. This relaxes the previously estimated
laser power requirements for upcoming facilities \cite{ELI} by a
factor of $3$ which is enormous in costs if ultra fast $\sim100J$
class lasers are considered.\\
The research leading to these results received funding from the
Deutsche Forschungsgemeinschaft within the program CRC/Transregio 18
and LASERLAB-EUROPE (grant agreement n$°$ $284464$, EC's Seventh
Framework Program). Computational resources was provided from the
ISC within the project MBU$15$.

\bibliography{Gold_arxiv}

\end{document}